\font\tenfrakturb=eufb10
\font\tenfraktur=eufm10
\font\tenmsbm=msbm10
\font\sevenfrakturb=eufb7
\font\sevenfraktur=eufm7
\font\sevenmsbm=msbm7
\font\fivefrakturb=eufb5
\font\fivefraktur=eufm5
\font\fivemsbm=msbm5
\newfam\bgothicfam
\newfam\gothicfam
\newfam\msbmfam
\textfont\bgothicfam = \tenfrakturb \scriptfont\bgothicfam=\sevenfrakturb
\scriptscriptfont\bgothicfam=\fivefrakturb
\textfont\gothicfam = \tenfraktur \scriptfont\gothicfam=\sevenfraktur
\scriptscriptfont\gothicfam=\fivefraktur
\textfont\msbmfam = \tenmsbm \scriptfont\msbmfam=\sevenmsbm
\scriptscriptfont\msbmfam=\fivemsbm

\def\Bbb{\tenmsbm\fam\msbmfam}

\catcode`@=11
\def\renewcounter#1{\@definecounter{#1}\@ifnextchar[{\@newctr{#1}}{}}

\documentstyle[twoside]{article}
\catcode`\@=11
\long\def\@makefntext#1{
\protect\noindent \hbox to 3.2pt {\hskip-.9pt  
$^{{\eightrm\@thefnmark}}$\hfil}#1\hfill} 

\def\@makefnmark{\hbox to 0pt{$^{\@thefnmark}$\hss}} 
	
\def\ps@myheadings{\let\@mkboth\@gobbletwo
\def\@oddhead{\hbox{}
\rightmark\hfil\eightrm\thepage}   
\def\@oddfoot{}\def\@evenhead{\eightrm\thepage\hfil
\leftmark\hbox{}}\def\@evenfoot{}
\def\sectionmark##1{}\def\subsectionmark##1{}}
\oddsidemargin=\evensidemargin
\newcounter{sectionc}\newcounter{subsectionc}\newcounter{subsubsectionc}
\renewcommand{\section}[1] {\vspace{12pt}\addtocounter{sectionc}{1} 
\setcounter{subsectionc}{0}\setcounter{subsubsectionc}{0}\noindent 
	{\tenbf\thesectionc. #1}\par\vspace{5pt}}
\renewcommand{\subsection}[1] {\vspace{12pt}\addtocounter{subsectionc}{1} 
	\setcounter{subsubsectionc}{0}\noindent 
	{\bf\thesectionc.\thesubsectionc. {\kern1pt \bfit #1}}\par\vspace{5pt}}
\renewcommand{\subsubsection}[1] {\vspace{12pt}\addtocounter{subsubsectionc}{1}
	\noindent{\tenrm\thesectionc.\thesubsectionc.\thesubsubsectionc.
	{\kern1pt \tenit #1}}\par\vspace{5pt}}
\newcommand{\nonumsection}[1] {\vspace{12pt}\noindent{\tenbf #1}
	\par\vspace{5pt}}
\newcounter{appendixc}
\newcounter{subappendixc}[appendixc]
\newcounter{subsubappendixc}[subappendixc]
\renewcommand{\thesubappendixc}{\Alph{appendixc}.\arabic{subappendixc}}
\renewcommand{\thesubsubappendixc}
	{\Alph{appendixc}.\arabic{subappendixc}.\arabic{subsubappendixc}}
\renewcommand{\appendix}[1] {\vspace{12pt}
        \refstepcounter{appendixc}
        \setcounter{figure}{0}
        \setcounter{table}{0}
        \setcounter{lemma}{0}
        \setcounter{theorem}{0}
        \setcounter{corollary}{0}
        \setcounter{definition}{0}
        \setcounter{equation}{0}
        \renewcommand{\thefigure}{\Alph{appendixc}.\arabic{figure}}
        \renewcommand{\thetable}{\Alph{appendixc}.\arabic{table}}
        \renewcommand{\theappendixc}{\Alph{appendixc}}
        \renewcommand{\thelemma}{\Alph{appendixc}.\arabic{lemma}}
        \renewcommand{\thetheorem}{\Alph{appendixc}.\arabic{theorem}}
        \renewcommand{\thedefinition}{\Alph{appendixc}.\arabic{definition}}
        \renewcommand{\thecorollary}{\Alph{appendixc}.\arabic{corollary}}
        \renewcommand{\theequation}{\Alph{appendixc}.\arabic{equation}}
        \noindent{\tenbf Appendix \theappendixc #1}\par\vspace{5pt}}
\newcommand{\subappendix}[1] {\vspace{12pt}
        \refstepcounter{subappendixc}
        \noindent{\bf Appendix \thesubappendixc. {\kern1pt \bfit #1}}
	\par\vspace{5pt}}
\newcommand{\subsubappendix}[1] {\vspace{12pt}
        \refstepcounter{subsubappendixc}
        \noindent{\rm Appendix \thesubsubappendixc. {\kern1pt \tenit #1}}
	\par\vspace{5pt}}
\newcommand{\textlineskip}{\baselineskip=13pt}
\newcommand{\smalllineskip}{\baselineskip=10pt}
\def\eightcirc{
\begin{picture}(0,0)
\put(4.4,1.8){\circle{6.5}}
\end{picture}}
\def\eightcopyright{\eightcirc\kern2.7pt\hbox{\eightrm c}} 
\newcommand{\copyrightheading}[1]
	{\vspace*{-2.5cm}\smalllineskip{\flushleft
	{\footnotesize Modern Physics Letters A, #1}\\
	{\footnotesize $\eightcopyright$\, World Scientific Publishing
	 Company}\\
         }}
\newcommand{\pub}[1]{{\begin{center}\footnotesize\smalllineskip 
	Received #1\\
	\end{center}
        }}

\def\abstracts#1#2#3{{
        \centering{\begin{minipage}{4.5in}\baselineskip=10pt\footnotesize
        \parindent=0pt #1\par 
        \parindent=15pt #2\par
        \parindent=15pt #3
        \end{minipage}}\par}} 

\newcommand{\bibit}{\nineit}
\newcommand{\bibbf}{\ninebf}
\renewenvironment{thebibliography}[1]
         {\frenchspacing
         \ninerm\baselineskip=11pt
         \begin{list}{\arabic{enumi}.}
         {\usecounter{enumi}\setlength{\parsep}{0pt}     
         \setlength{\leftmargin 12.7pt}{\rightmargin 0pt} 
         \setlength{\itemsep}{0pt} \settowidth
         {\labelwidth}{#1.}\sloppy}}{\end{list}}
\newcounter{itemlistc}
\newcounter{romanlistc}
\newcounter{alphlistc}
\newcounter{arabiclistc}

\newcommand{\fcaption}[1]{
         \refstepcounter{figure}
         \setbox\@tempboxa = \hbox{\footnotesize Fig.~\thefigure. #1}
         \ifdim \wd\@tempboxa > 5in
           {\begin{center}
         \parbox{5in}{\footnotesize\smalllineskip Fig.~\thefigure. #1}
            \end{center}}
        \else
             {\begin{center}
             {\footnotesize Fig.~\thefigure. #1}
              \end{center}}
        \fi}
\newcommand{\tcaption}[1]{
        \refstepcounter{table}
        \setbox\@tempboxa = \hbox{\footnotesize Table~\thetable. #1}
        \ifdim \wd\@tempboxa > 5in
           {\begin{center}
        \parbox{5in}{\footnotesize\smalllineskip Table~\thetable. #1}
            \end{center}}
        \else
             {\begin{center}
             {\footnotesize Table~\thetable. #1}
              \end{center}}
        \fi}
\def\@citex[#1]#2{\if@filesw\immediate\write\@auxout
        {\string\citation{#2}}\fi
\def\@citea{}\@cite{\@for\@citeb:=#2\do
        {\@citea\def\@citea{,}\@ifundefined
        {b@\@citeb}{{\bf ?}\@warning
        {Citation `\@citeb' on page \thepage \space undefined}}
        {\csname b@\@citeb\endcsname}}}{#1}}
\newif\if@cghi
\def\cite{\@cghitrue\@ifnextchar [{\@tempswatrue
        \@citex}{\@tempswafalse\@citex[]}}
\def\citelow{\@cghifalse\@ifnextchar [{\@tempswatrue
        \@citex}{\@tempswafalse\@citex[]}}
\def\@cite#1#2{{$\null^{#1}$\if@tempswa\typeout
        {IJCGA warning: optional citation argument 
        ignored: `#2'} \fi}}

\def\pmb#1{\setbox0=\hbox{#1}
        \kern-.025em\copy0\kern-\wd0
        \kern.05em\copy0\kern-\wd0
        \kern-.025em\raise.0433em\box0}

\def\fnt#1#2{\footnotetext{\kern-.3em
        {$^{\mbox{\scriptsize #1}}$}{#2}}}
\def\fpage#1{\begingroup
\voffset=.3in
\thispagestyle{empty}\begin{table}[b]\centerline{\footnotesize #1}
       \end{table}\endgroup}
\def\runninghead#1#2{\pagestyle{myheadings}
\markboth{{\protect\footnotesize\it{\quad #1}}\hfill}
{\hfill{\protect\footnotesize\it{#2\quad}}}}
\font\tenrm=cmr10
\font\tenit=cmti10 
\font\tenbf=cmbx10
\font\bfit=cmbxti10 at 10pt
\font\ninerm=cmr9
\font\nineit=cmti9
\font\ninebf=cmbx9
\font\eightrm=cmr8

\def\qed{\hbox{${\vcenter{\vbox{  
   \hrule height 0.4pt\hbox{\vrule width 0.4pt height 6pt
   \kern5pt\vrule width 0.4pt}\hrule height 0.4pt}}}$}}

\begin{document}
\runninghead{Yu. P. Goncharov \& E. A. Choban}
{ Dirac equation in the confining SU(3)-Yang-Mills field and 
the relativistic effects in quarkonia spectra}
\normalsize\textlineskip
\thispagestyle{empty}
\setcounter{page}{1}
\copyrightheading{Vol. 18, No. 24 (2003) 1661-1671}
\vspace*{0.88truein}
\fpage{1}
\centerline{\bf DIRAC EQUATION IN THE CONFINING SU(3)-YANG-MILLS FIELD}
\vspace*{0.035truein}
\centerline{\bf AND THE RELATIVISTIC EFFECTS IN QUARKONIA SPECTRA}
\vspace*{0.37truein}
\centerline{\footnotesize YU. P. GONCHAROV}
\vspace*{0.015truein}
\centerline{\footnotesize\it Theoretical Group,
Experimental Physics Department, State Polytechnical University}
\baselineskip=10pt
\centerline{\footnotesize\it Sankt-Petersburg 195251, Russia}
\vspace*{10pt}
\centerline{\footnotesize E. A. CHOBAN\footnote{Deceased.}}
\vspace*{0.015truein}
\centerline{\footnotesize\it Theoretical Physics Department,
State Polytechnical University}
\baselineskip=10pt
\centerline{\footnotesize\it Sankt-Petersburg 195251, Russia}
\vspace*{10pt}
\vspace*{0.225truein}
\pub{5 January 2003}
\vspace*{0.21truein}
\abstracts{
The recently obtained solutions of Dirac equation in the confining
SU(3)-Yang-Mills field in Minkowski spacetime are applied to describe
the energy spectra of quarkonia (charmonium and bottomonium). The nonrelativistic 
limit is considered for the relativistic effects to be estimated in a self-consistent 
way and it is shown that the given effects are extremely important for both 
the energy spectra and the confinement mechanism.
}{}{}
\vspace*{1pt}\textlineskip 
\section{Introductory remarks} 
\vspace*{-0.5pt}
\noindent
   Theory of quarkonium ranks high within the hadron physics as the one of
central sources of information about the quark interaction. Referring for
more details to the recent up-to-date review,\cite{Grin00} it should be
noted here that at present some generally accepted relativistic model of
quarkonium is absent. The description of quarkonium is actually implemented
by nonrelativistic manner (on the basis of the Schr{\"o}dinger equation) and
then one tries to include relativistic corrections in one or another way.
Such an inclusion is not single-valued and varies in dependence of the point
of view for different authors (see, e. g. Ref.\cite{Du00} and references
therein). It would be more consistent, to our mind, building a primordially
relativistic model so that one can then pass on to the nonrelativistic one by
the standard limiting transition and, thus, to estimate the relativistic
effects in a self-consistent way.

 As follows from the main principles of quantum chromodynamics (QCD),
the suitable relativistic models for description of relativistic bound states
of quarkonium should consist in considering the solutions of Dirac equation
in a SU(3)-Yang-Mills field representing gluonic field. The latter should be
the so-called confining solution of the corresponding Yang-Mills equations and
should model the quark confinement. Such solutions are usually supposed
to contain at least one
component of the mentioned SU(3)-field linear in $r$, the distance between
quarks. Recently in Ref.\cite{Gon01} a number of such solutions has been
obtained and the corresponding spectrum of the Dirac equation describing the
relativistic bound states in this confining SU(3)-Yang-Mills field
has been analysed. In this note we should like to apply the results of
Ref.\cite{Gon01} to description of the charmonium and bottomonium spectra.
We here solve the inverse problem, i. e. we define the confining gluonic field
components in the covariant description (SU(3)-connection) for charmonium and
bottomonium employing the experimental data on the mentioned spectra.\cite{pdg} 
As a result, we shall not use any nonrelativistic potentials modelling 
confinement, for example, of the harmonic oscillator or funnel type, in 
particular, because the latter ones do not satisfy the Yang-Mills equation
while the SU(3)-gluonic field used by us does. In our case the approach is
relativistic from the very outset and our considerations are essentially 
nonperturbative since we shall not use any expansions in the coupling constant
$g$ or in any other parameters.

Further we shall deal with the metric of
the flat Minkowski spacetime $M$ that
we write down (using the ordinary set of local spherical coordinates
$r,\vartheta,\varphi$ for spatial part) in the form
$$ds^2=g_{\mu\nu}dx^\mu\otimes dx^\nu\equiv
dt^2-dr^2-r^2(d\vartheta^2+\sin^2\vartheta d\varphi^2)\>. \eqno(1)$$
Besides we have $|\delta|=|\det(g_{\mu\nu})|=(r^2\sin\vartheta)^2$
and $0\leq r<\infty$, $0\leq\vartheta<\pi$,
$0\leq\varphi<2\pi$.

  Throughout the paper we employ the system of units with $\hbar=c=1$,
unless explicitly stated otherwise.
Finally, we shall denote $L_2(F)$ the set of the modulo square integrable
complex functions on any manifold $F$ furnished with an integration measure
while $L^n_2(F)$ will be the $n$-fold direct product of $L_2(F)$
endowed with the obvious scalar product.

\section{Preliminary considerations}

\subsection{Dirac equation}

  To formulate the results of Ref.\cite{Gon01} needed to us here, let us
notice that the relativistic wave function of quarkonium can be chosen
in the form
$$\psi=\pmatrix{\psi_1\cr\psi_2\cr\psi_3\cr}$$
with the four-dimensional spinors $\psi_j$ representing $j$-th colour
component of quarkonium. The corresponding Dirac equation
for $\psi$ may look as follows
$${\cal D}\psi=\mu_0\psi,\>\eqno(2)$$
where $\mu_0$ is a mass parameter
while the coordinate $r$ makes sense of the distance between quarks.

From general considerations the explicit form of
the operator ${\cal D}$ in local coordinates $x^\mu$ on Minkowski
manifold can be written as follows
$${\cal D}=i(\gamma^e\otimes I_3)E_e^\mu(\partial_\mu\otimes I_3
-\frac{1}{2}\omega_{\mu ab}\gamma^a\gamma^b\otimes I_3-igA_\mu),
\>a < b ,\>\eqno(3)$$
where $A=A_\mu dx^\mu$, $A_\mu=A^c_\mu T_c$ is a SU(3)-connection in the
(trivial) bundle $\xi$ over Minkowski spacetime, $I_3$ is the unit matrix
$3\times3$, the matrices $T_c$ form
a basis of the Lie algebra of SU(3) in 3-dimensional space (we consider
$T_a$ hermitean which is acceptable in physics), $c=1,...,8$, $\otimes$ here
means tensorial product of matrices, $g$ is a gauge coupling constant.
Further, the forms
$\omega_{ab}=\omega_{\mu ab}dx^\mu$ obey the Cartan structure equations
$de^a=\omega^a_{\ b}\wedge e^b$ with exterior derivative $d$, while the
orthonormal basis $e^a=e^a_\mu dx^\mu$ in cotangent bundle and
dual basis $E_a=E^\mu_a\partial_\mu$ in tangent bundle are connected by the
relations $e^a(E_b)=\delta^a_b$. At last, matrices $\gamma^a$ represent
the Clifford algebra of
the corresponding quadratic form $Q_{1,3}=x_0^2-x_1^2-x_2^2-x_3^2$
in ${\Bbb C}^{2}$.
For this we take the following choice for $\gamma^a$
$$\gamma^0=\pmatrix{1&0\cr 0&-1\cr}\,,
\gamma^b=\pmatrix{0&\sigma_b\cr-\sigma_b&0\cr}\,,
b= 1,2,3\>, \eqno(4)$$
where $\sigma_b$ denote the ordinary Pauli matrices.
It should be noted that, in lorentzian case, Greek indices $\mu,\nu,...$
are raised and lowered with $g_{\mu\nu}$ of (1) or its inverse $g^{\mu\nu}$
and Latin indices $a,b,...$ are raised and lowered by
$\eta_{ab}=\eta^{ab}$= diag(1,-1,-1,-1),
so that $e^a_\mu e^b_\nu g^{\mu\nu}=\eta^{ab}$,
$E^\mu_aE^\nu_bg_{\mu\nu}=\eta_{ab}$ and so on.

We can concretize the Dirac equation (2) for
$\psi$ in the case of metric (1).
Namely, we can put $e^0=dt$, $e^1=dr$,
$e^2=rd\vartheta$, $e^3=r\sin{\vartheta}d\varphi$ and, accordingly,
$E_0=\partial_t$, $E_1=\partial_r$,
$E_2=\partial_\vartheta/r$, $E_3=\partial_\varphi/(r\sin{\vartheta})$.
This entails
$$\omega_{12}=-d\vartheta,
\omega_{13}=-\sin{\vartheta}d\varphi,
\omega_{23}=-\cos{\vartheta}d\varphi.\>\eqno(5)$$
As for the connection $A_\mu$ in bundle $\xi$ then the suitable one should be
the confining solution of the Yang-Mills equations
$$dF=F\wedge A - A\wedge F \>, \eqno(6)$$
$$d\ast F= \ast F\wedge A - A\wedge\ast F \>\eqno(7)$$
with the exterior differential $d=\partial_t dt+\partial_r dr+
\partial_\vartheta d\vartheta+\partial_\varphi d\varphi$ in coordinates
$t,r,\vartheta,\varphi$ while the curvature matrix (field strentgh)
for $\xi$-bundle is $F=dA+A\wedge A$ and $\ast$ means the Hodge star
operator conforming to metric (1).
It is clear that (6) is identically satisfied --- this
is just the Bianchi identity holding true for any connection
so that it is necessary to solve only the equations (7).

\subsection{SU(3)-confining connection}
 In Ref.\cite{Gon01} the black hole physics techniques from
Refs.\cite{Gon678} was used
to find a set of the
confining solutions of Eq. (7).  For the aims of the given paper we need
one such a solution of Ref.\cite{Gon01} Let us adduce it here
putting $T_c=\lambda_c$,
where $\lambda_c$ are the Gell-Mann matrices (whose explicit form can be
found in Refs.\cite{Gon678}) Then the solution in question is the following
one
$$ A^3_t+\frac{1}{\sqrt{3}}A^8_t =-\frac{a_1}{r}+A_1 \>,
 -A^3_t+\frac{1}{\sqrt{3}}A^8_t=\frac{a_1+a_2}{r}-(A_1+A_2)\>,
-\frac{2}{\sqrt{3}}A^8_t=-\frac{a_2}{r}+A_2\>, $$
$$ A^3_\varphi+\frac{1}{\sqrt{3}}A^8_\varphi =b_1r+B_1 \>,
 -A^3_\varphi+\frac{1}{\sqrt{3}}A^8_\varphi=-(b_1+b_2)r-(B_1+B_2)\>,
-\frac{2}{\sqrt{3}}A^8_\varphi=b_2r+B_2\> \eqno(8)$$
with all other $A^c_\mu=0$, where real constants $a_j, A_j, b_j, B_j$
parametrize the solution, and we wrote down
the solution in the combinations that are just
needed to insert into (2). As is not complicated to see, the solution is
a configuration describing the electric Coulomb-like colour field (components
$A_t$) and the magnetic colour field linear in $r$ (components $A_\varphi$).
Also it is easy to check that the given solution satisfy the Lorentz gauge
condition that can be
written in the form ${\rm div}(A)=0$, where the divergence of the Lie algebra
valued 1-form $A=A^c_\mu T_cdx^\mu$ is defined by the relation
$${\rm div}(A)=\frac{1}{\sqrt{|\delta|}}\partial_\mu(\sqrt{|\delta|}g^{\mu\nu}
A_\nu)\>.\eqno(9)$$

\subsection{Dirac equation spectrum and wave functions}

 As was shown in Ref.\cite{Gon01}, after inserting the above confining
solution into Eq. (2), it admits the solutions of the form
$$\psi_j=e^{i\omega_j t}r^{-1}\pmatrix{F_{j1}(r)\Phi_j(\vartheta,\varphi)\cr\
F_{j2}(r)\sigma_1\Phi_j(\vartheta,\varphi)}\>,j=1,2,3\eqno(10)$$
with the 2D eigenspinor $\Phi_j=\pmatrix{\Phi_{j1}\cr\Phi_{j2}}$ of the
euclidean Dirac operator on the unit sphere ${\Bbb S}^2$.
The explicit form of $\Phi_j$ is not needed here and
can be found in
Refs.\cite{Gon99} For the purpose of the present paper it is sufficient
to know that spinors $\Phi_j$ can be subject to
the normalization condition
$$\int\limits_0^\pi\,\int\limits_0^{2\pi}(|\Phi_{j1}|^2+|\Phi_{j2}|^2)
\sin\vartheta d\vartheta d\varphi=1\> , \eqno(11)$$
i. e., they form an orthonormal basis in $L_2^2({\Bbb S}^2)$.

The energy spectrum $\varepsilon$ of quarkonium is given by the
relation $\varepsilon=\omega_1+\omega_2+\omega_3$ with
$$\omega_1=\omega_1(n_1,l_1,\lambda_1)=
\frac{-\Lambda_1 g^2a_1b_1+(n_1+\alpha_1)
\sqrt{(n_1^2+2n_1\alpha_1+\Lambda_1^2)\mu_0^2+g^2b_1^2(n_1^2+2n_1\alpha_1)}}
{n_1^2+2n_1\alpha_1+\Lambda_1^2}\>,\eqno(12)$$
$$\omega_2=\omega_2(n_2,l_2,\lambda_2)=$$
$$\frac{-\Lambda_2 g^2(a_1+a_2)(b_1+b_2)-(n_2+\alpha_2)
\sqrt{(n_2^2+2n_2\alpha_2+\Lambda_2^2)\mu_0^2+g^2(b_1+b_2)^2
(n_2^2+2n_2\alpha_2)}}
{n_2^2+2n_2\alpha_2+\Lambda_2^2}\>,\eqno(13)$$
$$\omega_3=\omega_3(n_3,l_3,\lambda_3)=
\frac{-\Lambda_3 g^2a_2b_2+(n_3+\alpha_3)
\sqrt{(n_3^2+2n_3\alpha_3+\Lambda_3^2)\mu_0^2+g^2b_2^2(n_3^2+2n_3\alpha_3)}}
{n_3^2+2n_3\alpha_3+\Lambda_3^2}\>,\eqno(14)$$
where
$\Lambda_1=\lambda_1-gB_1\>,\Lambda_2=\lambda_2+g(B_1+B_2)\>,
\Lambda_3=\lambda_3-gB_2\>,$
$n_j=0,1,2,...$, while $\lambda_j=\pm(l_j+1)$ are
the eigenvalues of euclidean Dirac operator
on unit sphere with $l_j=0,1,2,...$ Besides
$$\alpha_1=\sqrt{\Lambda_1^2-g^2a_1^2}\>,
\alpha_2=\sqrt{\Lambda_2^2-g^2(a_1+a_2)^2}\>,
\alpha_3=\sqrt{\Lambda_3^2-g^2a_2^2}\>.\eqno(15)$$

Further, the radial part of (10), for instance, for $\psi_1$-component, is
given at $n_1=0$ by
$$F_{11}=C_1Ar^{\alpha_1}e^{-\beta_1r}\left(1-
\frac{Y_1}{Z_1}\right),F_{12}=iC_1Br^{\alpha_1}e^{-\beta_1r}\left(1+
\frac{Y_1}{Z_1}\right),\eqno(16)$$
while at $n_1>0$ by
$$F_{11}=C_1Ar^{\alpha_1}e^{-\beta_1r}\left[\left(1-
\frac{Y_1}{Z_1}\right)L^{2\alpha_1}_{n_1}(r_1)+
\frac{AB}{Z_1}r_1L^{2\alpha_1+1}_{n_1-1}(r_1)\right],$$
$$F_{12}=iC_1Br^{\alpha_1}e^{-\beta_1r}\left[\left(1+
\frac{Y_1}{Z_1}\right)L^{2\alpha_1}_{n_1}(r_1)-
\frac{AB}{Z_1}r_1L^{2\alpha_1+1}_{n_1-1}(r_1)\right],\eqno(17)$$
with the Laguerre polynomials $L^\rho_{n_1}(r_1)$, $r_1=2\beta_1r$,
$\beta_1=\sqrt{\mu_0^2-(\omega_1-gA_1)^2+g^2b_1^2}$,
$A=gb_1+\beta_1$, $B=\mu_0+\omega_1-gA_1$,
$Y_1=[\alpha_1\beta_1- ga_1(\omega_1-gA_1)+g\alpha_1b_1]B+ g^2a_1b_1A$,
$Z_1=[(\lambda_1-gB_1)A+ga_1\mu_0)]B+ g^2a_1b_1A$.
Finally, $C_1$ is determined
from the normalization condition
$$\int_0^\infty(|F_{11}|^2+|F_{12}|^2)dr=\frac{1}{3}\>.\eqno(18)$$
Analogous relations will hold true for $\psi_{2,3}$, respectively,
by replacing
$a_1,A_1,b_1,B_1,\alpha_1 \to a_2,A_2,b_2,B_2,\alpha_3$ for $\psi_3$ and
$a_1,A_1,b_1,B_1,\alpha_1
\to -(a_1+a_2),-(A_1+A_2),-(b_1+b_2),-(B_1+B_2),\alpha_2$
for $\psi_2$ so that
$\beta_2=\sqrt{\mu_0^2-[\omega_2+g(A_1+A_2)]^2+g^2(b_1+b_2)^2}$,
$\beta_3=\sqrt{\mu_0^2-(\omega_3-gA_2)^2+g^2b_2^2}$.
Consequently, we shall gain that
$\psi_j\in L_2^{4}({\Bbb R}^3)$ at any $t\in{\Bbb R}$ and, as a result,
the solutions of (10) may describe relativistic bound states of quarkonium
with the energy spectrum (12)--(14).

\subsection{Nonrelativistic limit}
 Before apllying the above relations to a description of charmonium spectrum
let us adduce the nonrelativistic limits (i.e, at $c\to\infty$) for
the energies of (12)--(14). The common case is not needed to us in present
paper so we shall restrict ourselves to the case of $n_j=0,1$ and $l_j=0$.
Expanding $\omega_j$ in $x=\frac{g}{\hbar c}$, we get
$$\omega_1(0,0,\lambda_1)=-x\frac{ga_1b_1}{\lambda_1}+\mu_0c^2\left[1-
\frac{1}{2}\left(\frac{a_1}{\lambda_1}\right)^2x^2+O(x^3)\right]\>,$$
$$\omega_1(1,0,\lambda_1)=-x\frac{\lambda_1ga_1b_1}{(1+|\lambda_1|)^2}+
\mu_0c^2\left[1-
\frac{1}{8}\left(\frac{a_1}{\lambda_1}\right)^2x^2+O(x^3)\right]\>,\eqno(19)$$
which yields at $c\to\infty$ (putting $\hbar=c=1$ again)
$$\omega_1(0,0,\lambda_1)=
\mu_0\left[ 1-\frac{1}{2}\left(\frac{ga_1}{\lambda_1}\right)^2
\right]\>,
\omega_1(1,0,\lambda_1)=
\mu_0\left[ 1-\frac{1}{8}\left(\frac{ga_1}{\lambda_1}\right)^2
\right]\>.\eqno(20)$$
Analogously we shall have
$$\omega_2(0,0,\lambda_2)=-\mu_0\left[ 1-\frac{1}{2}\left(\frac{g(a_1+a_2)}
{\lambda_2}\right)^2\right]\>,
\omega_2(1,0,\lambda_2)=-\mu_0\left[ 1-\frac{1}{8}\left(\frac{g(a_1+a_2)}
{\lambda_2}\right)^2\right]\>,\eqno(21)$$
$$\omega_3(0,0,\lambda_3)=
\mu_0\left[ 1-\frac{1}{2}\left(\frac{ga_2}{\lambda_3}\right)^2
\right]\>,
\omega_3(1,0,\lambda_3)=
\mu_0\left[ 1-\frac{1}{8}\left(\frac{ga_2}{\lambda_3}\right)^2
\right]\>,\eqno(22)$$
where, of course, $\lambda_j=\pm1$ and $\lambda_j^2=1$.

\section{Relativistic spectrum of charmonium}
  Now we can adduce numerical results for constants parametrizing
the charmonium spectrum which are shown in Table 1.

\begin{table}[htbp]
\caption{Gauge coupling constant, mass parameter $\mu_0$ and
parameters of the confining SU(3)-connection for charmonium.}
\vskip 0.5truecm
\begin{tabular}{|c|c|c|c|c|c|c|c|}
\hline
$g$ & $\mu_0$, GeV & $a_1$  & $a_2$ & $b_1$, GeV & $b_2$, GeV & $B_1$ &
$B_2$ \\
\hline
0.618631 & 3.11409 & 0.0102202 & -0.126200 & -3.42927 & -3.91720 &
1.99496 & 2.10796 \\
\hline
\end{tabular}
\end{table}
As for parameters $A_{1,2}$ of solution (8), only the wave functions depend on
them while the spectrum does not and within the present paper we consider 
$A_1=A_2=0$.

With the constants of Table 1 the present-day levels of charmonium spectrum
were calculated with the help of (12)--(14) while their nonrelativistic
values with the aid of (20)--(22) according to the following combinations
(we use the notations of levels from Ref.\cite{pdg})
$$\eta_c(1S):
\varepsilon_1= \omega_1(0,0,-1)+\omega_2(0,0,-1)+\omega_3(0,0,-1)\>,$$
$$J/\psi(1S):
\varepsilon_2= \omega_1(0,0,-1)+\omega_2(0,0,1)+\omega_3(0,0,-1)\>,$$
$$\chi_{c0}(1P):
\varepsilon_3= \omega_1(0,0,-1)+\omega_2(0,0,-1)+\omega_3(0,0,1)\>,$$
$$\chi_{c1}(1P):
\varepsilon_4= \omega_1(1,0,-1)+\omega_2(0,0,1)+\omega_3(1,0,-1)\>,$$
$$\eta_{c}(1P):
\varepsilon_5= \omega_1(0,0,-1)+\omega_2(0,0,1)+\omega_3(0,0,1)\>,$$
$$\chi_{c2}(1P):
\varepsilon_6= \omega_1(1,0,1)+\omega_2(0,0,-1)+\omega_3(0,0,-1)\>,$$
$$\eta_c(2S):
\varepsilon_7= \omega_1(0,0,1)+\omega_2(0,0,-1)+\omega_3(1,0,1)\>,$$
$$\psi(2S):
\varepsilon_8= \omega_1(0,0,-1)+\omega_2(0,0,-1)+\omega_3(1,0,1)\>,$$
$$\psi(3770):
\varepsilon_9= \omega_1(1,0,-1)+\omega_2(0,0,-1)+\omega_3(0,0,1)\>,$$
$$\psi(4040):
\varepsilon_{10}= \omega_1(1,0,1)+\omega_2(0,0,-1)+\omega_3(0,0,1)\>,$$
$$\psi(4160):
\varepsilon_{11}= \omega_1(1,0,1)+\omega_2(0,0,1)+\omega_3(0,0,1)\>,$$
$$\psi(4415):
\varepsilon_{12}= \omega_1(1,0,1)+\omega_2(0,0,1)+\omega_3(1,0,1)\>.\eqno(23)$$

Table 2 contains experimental values of these levels (from Ref.\cite{pdg})
and our theoretical relativistic and nonrelativistic ones, and also the
contribution of relativistic effects in \%.

\begin{table}[htbp]
\caption{Experimental and theoretical charmonium levels}
\vskip 0.5truecm
\begin{tabular}{|c|c||c|c|c|}
\hline
$\varepsilon_j$ &  Experimental & Relativistic  & Nonrelativistic
GeV & Relativistic contribution\\
{} & Gev & GeV & GeV & \% \\
\hline
$\varepsilon_1$ & 2.97980  & 2.97980  & 3.11255 & 4.45516 \\
\hline
$\varepsilon_2$ & 3.09688  & 3.09688 & 3.11255 & 0.506165 \\
\hline
$\varepsilon_3$ & 3.41730  & 3.41730 & 3.11255 & 8.91779 \\
\hline
$\varepsilon_4$ & 3.51053  & 3.50702 & 3.11967 & 11.0450 \\
\hline
$\varepsilon_5$ & 3.52614  & 3.53438 & 3.11255 & 11.9350 \\
\hline
$\varepsilon_6$ & 3.55617  & 3.61609 & 3.11260 & 13.9237 \\
\hline
$\varepsilon_7$ & 3.59400  & 3.66485 & 3.11967 & 14.8758 \\
\hline
$\varepsilon_8$ & 3.68600  & 3.71725 & 3.11967 & 16.0759 \\
\hline
$\varepsilon_9$ & 3.76990  & 3.77773 & 3.11260 & 17.6066 \\
\hline
$\varepsilon_{10}$ & 4.04000  & 4.05359 & 3.11260 & 23.2138 \\
\hline
$\varepsilon_{11}$ & 4.16000  & 4.17067 & 3.11260 & 25.3694 \\
\hline
$\varepsilon_{12}$ & 4.41500  & 4.47062 & 3.11972 & 30.2174 \\
\hline
\end{tabular}
\end{table}

\section{Relativistic spectrum of bottomonium}
Numerical values for the corresponding constants parametrizing
the bottomonium spectrum which are shown in Table 3.
\begin{table}[htbp]
\caption{Gauge coupling constant, mass parameter $\mu_0$ and
parameters of the confining SU(3)-connection for bottomonium.}
\vskip 0.5truecm
\begin{tabular}{|c|c|c|c|c|c|c|c|}
\hline
$g$ & $\mu_0$, GeV & $a_1$  & $a_2$ & $b_1$, GeV & $b_2$, GeV & $B_1$ &
$B_2$ \\
\hline
0.172961 & 9.71992 & -0.313881 & -0.188848 & -21.1329 & 34.2764 &
1.48168 & 2.01632 \\
\hline
\end{tabular}
\end{table}
With the constants of Table 3 the present-day levels of bottomonium spectrum
were calculated with the help of (12)--(14) while their nonrelativistic
values with the aid of (20)--(22) according to the following combinations
(we use the notations of levels from Ref.\cite{pdg})
$$\Upsilon(1S):
\varepsilon_1= \omega_1(0,0,-1)+\omega_2(0,0,-1)+\omega_3(0,0,-1)\>,$$
$$\chi_{b0}(1P):
\varepsilon_2= \omega_1(0,0,-1)+\omega_2(0,0,1)+\omega_3(0,0,-1)\>,$$
$$\chi_{b1}(1P):
\varepsilon_3= \omega_1(0,0,-1)+\omega_2(0,0,-1)+\omega_3(0,0,1)\>,$$
$$\chi_{b2}(1P):
\varepsilon_4= \omega_1(1,0,1)+\omega_2(1,0,1)+\omega_3(0,0,-1)\>,$$
$$\Upsilon(2S):
\varepsilon_5= \omega_1(0,0,-1)+\omega_2(1,0,1)+\omega_3(0,0,1)\>,$$
$$\chi_{b0}(2P):
\varepsilon_6= \omega_1(1,0,1)+\omega_2(1,0,-1)+\omega_3(0,0,1)\>,$$
$$\chi_{b1}(2P):
\varepsilon_7= \omega_1(1,0,-1)+\omega_2(0,0,-1)+\omega_3(0,0,1)\>,$$
$$\chi_{b2}(2P):
\varepsilon_8= \omega_1(0,0,1)+\omega_2(0,0,-1)+\omega_3(1,0,-1)\>,$$
$$\Upsilon(3S):
\varepsilon_9= \omega_1(1,0,1)+\omega_2(1,0,1)+\omega_3(0,0,1)\>,$$
$$\Upsilon(4S):
\varepsilon_{10}= \omega_1(0,0,1)+\omega_2(1,0,-1)+\omega_3(1,0,1)\>,$$
$$\Upsilon(10860):
\varepsilon_{11}= \omega_1(0,0,-1)+\omega_2(1,0,1)+\omega_3(1,0,-1)\>,$$
$$\Upsilon(11020):
\varepsilon_{12}= \omega_1(0,0,1)+\omega_2(0,0,1)+\omega_3(1,0,1)\>.\eqno(24)$$

Table 4 contains experimental values of these levels (from Ref.\cite{pdg})
and our theoretical relativistic and nonrelativistic ones, and also the
contribution of relativistic effects in \%.
\begin{table}[htbp]
\caption{Experimental and theoretical bottomonium levels}
\vskip 0.5truecm
\begin{tabular}{|c|c||c|c|c|}
\hline
$\varepsilon_j$ &  Experimental & Relativistic  & Nonrelativistic
 & Relativistic contribution\\
{} & Gev & GeV & GeV & \% \\
\hline
$\varepsilon_1$ & 9.46037  & 9.46037  & 9.73716 & 2.92574 \\
\hline
$\varepsilon_2$ & 9.8598  & 9.8598 & 9.73716 & 1.24391 \\
\hline
$\varepsilon_3$ & 9.8919  & 9.8919 & 9.73716 & 1.56436 \\
\hline
$\varepsilon_4$ & 9.9132  & 9.95542 & 9.72034 & 2.36137 \\
\hline
$\varepsilon_5$ & 10.02330  & 10.0427 & 9.70960 & 3.31702 \\
\hline
$\varepsilon_6$ & 10.2321  & 10.2359 & 9.72034 & 5.03641 \\
\hline
$\varepsilon_7$ & 10.2552  & 10.2521 & 9.74790 & 4.91837 \\
\hline
$\varepsilon_8$ & 10.2685  & 10.2611 & 9.74104 & 5.06841 \\
\hline
$\varepsilon_9$ & 10.3553  & 10.3870 & 9.72034 & 6.41781 \\
\hline
$\varepsilon_{10}$ & 10.5800  & 10.6315 & 9.71349 & 8.64370 \\
\hline
$\varepsilon_{11}$ & 10.8650  & 10.8536 & 9.71349 & 10.5041 \\
\hline
$\varepsilon_{12}$ & 11.019  & 11.0312 & 9.74104 & 11.6955 \\
\hline
\end{tabular}
\end{table}
\newpage
\section{Physical interpretation}
\subsection{Role of the colour magnetic field}
 The results obtained allow us to draw a number of conclusions. As is seen
from Tables 2,4, relativistic values are in good agreement with experimental
ones while nonrelativistic ones are not. The contribution of relativistic
effects can amount to tens per cent and they cannot be considered as
small. The physical reason of it is quite clear. Really, we have seen in
nonrelativistic limit [see the relations (19)--(22)] that parameters
$b_{1,2}, B_{1,2}$ [see Eq. (8)] of linear interaction between quarks vanish
under this limit and
nonrelativistic spectrum is independent of them and is practically getting
the pure Coulomb one. As a consequence, the picture of linear confinement for
quarks should be considered as essentially relativistic one while
the nonrelativistic limit is only a rather crude approximation. In fact, as 
follows from exact solutions of SU(3)-Yang--Mills equations of (8), the linear
interaction between quarks is connected with colour magnetic field that
dies out in the nonrelativistic limit, i.e. for static quarks. Only for
the moving rapidly enough quarks the above field will appear and generate
linear confinement between them. So the spectrum will depend on both
the static Coulomb colour electric field and the dynamical colour magnetic
field responsible for the linear confinement for quarks which is just confirmed
by the relations (12)--(14). For bottomonium the mentioned effects are smaller
than for charmonium which well corresponds to physics in question -- in the
former case quarks are more massive and relativistic effects should be smaller.
Also one can say in other words that colour magnetic
field splits the primordially nonrelativistic spectrum into some fine structure
as, for example, the Seemann effect does in atomic physics. But, unlike the 
latter case, for quarkonia these corrections are not small. 

\subsection{Specification of the wave function form}
One can notice that
the form of wave functions (16)--(17) permits to consider, for instance,
the quantity $1/\beta_1$ to be a characteristic size of quarkonium. Under the
circumstances, if calculating $1/\beta_1$ in both relativistic and
nonrelativistic cases then one can obtain, for example, for the charmonium
$\eta_c(2S)$-level that relativistic size will be of order
$r\sim1/\beta_1\sim0.938854\cdot10^{-14}$ cm while nonrelativistic one
(i.e. if calculating $\beta_1$ at $b_1=0$) would be of order
$r_0\sim0.101539\cdot10^{-11}$ cm, i. e., $r_0/r\sim108>>1$. 
Analogously, e. g., for the bottomonium $\Upsilon(1S)$-level the conforming
quantities will be: $r\sim0.618867\cdot10^{-14}$ cm,
$r_0\sim0.379148\cdot10^{-13}$ cm,$r_0/r\sim6>1$. This additionally
points out the importance of relativistic effects for confinement.

\subsection{Comparison with the potential approach}
One should say a few words concerning the nonrelativistic potential models
often used in quarkonium theory.\cite{GM} The potentials between quarks here
are usually modelled by those of harmonic oscillator or of funnel type (i. e.
of the form $\alpha/r+\beta r$ with some constants $\alpha$ and $\beta$).
It is clear, however, that from the QCD point of view the interaction between
quarks should be described by the whole SU(3)-connection $A_\mu=A^c_\mu T_c$,
genuinely relativistic object, the nonrelativistic potential being only some
component of $A^c_t$ surviving in the nonrelativistic limit at $c\to\infty$.
As is easy to show, however, the SU(3)-connection of form $A^c_t=Br^\gamma$,
where $B$ is a constant, may be solution of the Yang-Mills equations (7) only
at $\gamma=-1$, i. e. in the Coulomb-like case. As a result, the potentials
employed in nonrelativistic approaches do not obey the Yang-Mills equations.
The latter ones are essentially relativistic and, as we have seen, the
components linear in $r$ of the whole $A_\mu$ are different from $A_t$ and
related with colour magnetic field vanishing in the nonrelativistic limit.
That is why the nonrelativistic potential approach seems to be inconsistent.
Our approach uses only the exact solutions of the Yang-Mills equations as well
as in atomic physics the interaction among particles (e. g. the electric
Coulomb one) is always the exact solution of the Maxwell equations (the
particular case of the Yang-Mills equations).

\section{Concluding remarks}
The calculations of the present paper can be extended.
Indeed we have the explicit form (10) for the relativistic
wave functions of quarkonium that may be applied to analysis of the quarkonium
radiative decays and electromagnetic transitions. Unfortunately in connection
with death of E. Choban the work in this direction is not yet completed but
the first author hopes to discuss the mentioned questions elsewhere.

\section{Acknowlegments}
    The work of Goncharov was supported in part by the Russian
Foundation for Basic Research (grant no. 01-02-17157).
\vskip1cm
\centerline{\bf To the memory of E. A. Choban}
In the process of working at the present paper it happened big misfortune
--late in May 2002 E. Choban suddenly died so the paper has been finished
without him. I tried to keep everything that we could discuss together. Let
the given paper be the last tribute to the remarkable man and enthusiast of
hadronic physics who was Enver Choban.

\nonumsection{References}

\end{document}